\documentclass[letterpaper,10pt,conference]{ieeeconf}
\pdfoutput=1

\usepackage{color}
\usepackage{times}
\usepackage{flushend}
\usepackage[dvipsnames]{xcolor}
\usepackage{amsmath, mathtools}
\usepackage{amssymb}
\usepackage{graphicx}
\usepackage{epsfig}
\usepackage{bm}
\usepackage{url}
\usepackage[font=footnotesize, skip=4pt]{caption}
\usepackage{subfig}
\usepackage{amsthm}

\newtheorem{lemma}{Lemma}
\newtheorem{theorem}{Theorem}

\newtheorem{proposition}{Proposition}
\newtheorem{assumption}{Assumption}

\IEEEoverridecommandlockouts
\usepackage{tikz}
\usetikzlibrary{shapes,arrows}
\usetikzlibrary{bayesnet}
\usetikzlibrary{positioning}
\tikzstyle{status} = [rectangle, draw=black, text centered, anchor=north, text=black, minimum width=9em, minimum height=3em, node distance=6ex and 7em]
\tikzstyle{line} = [draw,thick,-latex]
\tikzstyle{transition} = [font=\small]

\makeatletter
\newcommand{\revisionhistory}[1]{%
\@ifundefined{showrevisionhistory}{\relax}{%
{#1}%
}%
}
\makeatother

\begin{document}


\title{\LARGE \bf Hedging Strategies for Load-Serving Entities\\in Wholesale Electricity Markets}
\author{Datong P. Zhou$^{\ast\dagger}$, Munther A. Dahleh$^\dagger$, and Claire J. Tomlin$^\star$
\thanks{$^\ast$Department of Mechanical Engineering, University of California, Berkeley, USA. {\tt\footnotesize datong.zhou@berkeley.edu}}
\thanks{$^\dagger$Laboratory for Information and Decision Systems, MIT, Cambridge, USA.
{\tt\footnotesize [datong,dahleh]@mit.edu}}
\thanks{$^\star$Department of Electrical Engineering and Computer Sciences, University of California, Berkeley, USA.
{\tt\footnotesize tomlin@eecs.berkeley.edu}}%
\thanks{This work has been supported in part by the National Science Foundation under CPS:FORCES (CNS-1239166) and CEC Grant 15-083.}
}

\maketitle
\thispagestyle{empty}
\pagestyle{empty}

\begin{abstract}
Load-serving entities which procure electricity from the wholesale electricity market to service end-users face significant quantity and price risks due to the volatile nature of electricity demand and quasi-fixed residential tariffs at which electricity is sold. This paper investigates strategies for load serving entities to hedge against such price risks. Specifically, we compute profit-maximizing portfolios of forward contract and call options as a function of uncertain aggregate user demand and wholesale electricity prices. We compare the profit to the case of Demand Response, where users are offered monetary incentives to temporarily reduce their consumption during periods of supply shortages. Using smart meter data of residential customers in California, we simulate optimal portfolios and derive conditions under which Demand Response outperforms call options and forward contracts. Our analysis suggests that Demand Response becomes more competitive as wholesale electricity prices increase.
\end{abstract}


%
%


\section{Introduction}
\label{sec:Introduction}
Historically, electricity was supplied by vertically integrated entities which maintained full functional control over the entire supply chain, including generation, transmission, and distribution assets. This static structure constituted an impediment for new energy providers on both the supply and retail end to participate in the energy market. In the United States, the Federal Energy Regulatory Commission issued Orders 888 and 889 in April 1996 to remove such barriers of entry in an attempt to promote competition and market efficiency \cite{FERC888, FERC889}. The result of this market design process was a combination between a central electricity pool operating day-ahead, overseen by Independent System Operators (ISOs), and bilateral trading between generating companies and electric utilities, which supplanted the traditional, vertically integrated entities.

As a consequence of the restructuring process, generators and utilities in the electricity market started facing price and quantity risks ensuing from the inelasticity of user demand, the steep supply curve due to the slowly changing nature of power plants' output adjustment, and prohibitive cost of energy storage. These factors allow small increases or decreases of demand to result in a price boom or bust, respectively. Furthermore, despite the fact that the economic consensus calls for passing along varying electricity prices to end-users in order to increase economic efficiency \cite{Borenstein:2002ab, Borenstein:2005aa, Borenstein:2005ab}, policymakers have retained quasi-fixed electricity tariffs, e.g. Time-of-Use pricing. In conjunction with the obligation of utilities to service end-users with electricity at all times, risks associated with sudden price spikes are borne by the utility. This market situation has resulted in several crises. For instance, unseasonably warm climate in the summer of 2000 resulted in California's wholesale electricity prices to rise to average prices of more than 140 USD/MWh, leading to the bankruptcy of Pacific Gas \& Electric, California's largest utility, and high profits of electricity generators \cite{Borenstein:2002aa}. Similar crises occurred in Texas (2004) and in the Midwestern United States (1998).

These crises resulted in the following notable developments. Firstly, electric utilities and generating companies started to hedge against price fluctuations through contracts on different scales of time, ranging from short-term forward contracts to long-term contracts, thereby locking in a fixed price and quantity to be delivered over a contractually specified period of time. Secondly, Demand-Side Management (DSM), which aims to affect consumer behavior during periods of peak demand, emerged as a viable tool to partially relay price risks to end-users. For instance, companies like \texttt{OPOWER} provide Demand Response (DR) services to utilities, allowing them to offer monetary rewards to end-users in exchange for a reduction in electricity consumption during hours of peak demand.

Motivated by these shortcomings, a large body of research, particularly in operations research, has studied optimal hedging contracts, most often from the utility perspective, including \cite{Oum:2006aa, Oum:2009aa}, where the authors construct an optimal one-step hedging portfolio with standard power options, or \cite{Hatami:2009aa}, which finds an optimal energy procurement policy with stochastic programming over a specified period. \cite{Borenstein:2006aa} analyzes hedging instruments against price volatility for industrial customers. \cite{Wolak:2001aa} investigates hedging strategies for electricity generators.

While there exists a large body of literature on operational and algorithmic aspects of DR (e.g. load scheduling and shifting \cite{Mohsenian-Rad:2010aa, Li:2011aa, Aalami:2010aa}), significantly less research has focused on the role of DR programs as an alternative way of hedging. Notable examples are \cite{Deng:2009aa}, where the authors discuss interruptible service contracts, and \cite{Sezgen:2007aa}, which estimates the economic value of DR programs for commerical customers by adapting models used to value energy options. To the best of our knowledge, no significant research has investigated the option value of residential DR programs. To close this gap, we derive a stylized model for the utility's profit under such DR schemes and determine its optimal profit. The methodology we use is closest in spirit with \cite{Bitar:2012aa}, where the authors determine the optimal bidding volume of wind generators in a conventional energy market. We compare the profit under Demand Response to the case of forward contracts and call options by incorporating the conditional value at risk \cite{Rockafellar:2002aa} measure. Using smart meter data of residential customers in California, we find that DR can yield higher expected profits than under forward contracts and call options, especially in the presence of high wholesale electricity prices.

The remainder of this paper is organized as follows: In Section \ref{sec:Market_Participants}, we describe the interactions between the participants in energy markets. Section \ref{sec:Options} introduces forward contracts, options, and Demand Response as hedging instruments for the Demand Response Provider. The effect of uncertainty in the user demand on the expected profit of the Demand Response Provider is investigated in Section \ref{sec:Uncertainty}. We compute optimal, profit-maximizing portfolios for load-serving entities in Section \ref{sec:best_option} and simulate decision boundaries between them in Section \ref{sec:simulations}. Section \ref{sec:Conclusion} concludes. All proofs are relegated to the Appendix.

\subsection*{Notation}
Let $\mathbb{E}[\hspace{0.025cm}\cdot\hspace{0.025cm}]$ denote the expectation of a random variable. Let $\left[\hspace{0.025cm}\cdot\hspace{0.025cm}\right]_+$ denote the hinge function, i.e. $[\hspace{0.015cm} x \hspace{0.015cm}]_+ = \max(0,x)$.

%
%


\section{Market Participants}
\label{sec:Market_Participants}

\begin{figure}[h]
\centering
\begin{tikzpicture}
\node [status, thick, align=left, fill={rgb:black,0.5;white,3}] (Gens) {\textbf{Generators}};
\node [status, dashed, below=3em of Gens, fill={rgb:black,0.5;white,3}] (WSM) {Wholesale Market};
\node [status, thick,below=3em of WSM, fill={rgb:black,0.5;white,3}] (Ut) {\textbf{Utility}};
\node [status, thick,right=2em of WSM, fill={rgb:black,0.5;white,3}] (EU) {\textbf{End Users}};
\path [line, line width=1.1] ([yshift=0.8em]Ut.east) -- ++(1.9,0) -- ++(0.0,1.0) -- ([xshift=-1.1em]EU.south) node [transition,pos=0.2,left] {$\boldsymbol{\lambda_f, r}$};
\path [line, line width=1.1] ([xshift=1.1em]EU.south) -- ++(0,-1.87) -- ++(-1.9,0.0) -- ([yshift=-0.8em]Ut.east) node [transition,pos=-3.2,above] {$\boldsymbol{d, h(r)}$};

\path [line, line width=1.1] ([xshift=-1.3em]WSM.south) -- ++(0,-0.67) -- ++ (0,0) -- ([xshift=-1.3em]Ut.north) node [transition,pos=0.2,left] {$\boldsymbol{[d-\bar{q}]_{+}}$};
\path [line, line width=1.1] ([xshift=1.3em]Ut.north) -- ++(0,0.67) -- ++ (0,0) -- ([xshift=1.3em]WSM.south) node [transition,pos=0.2,right] {$\boldsymbol{\lambda_s}$};

\path [line] ([xshift=-1.3em]Gens.south) -- ++(0,-0.67) -- ++ (0,0) -- ([xshift=-1.3em]WSM.north) [dashed] node [transition,pos=0.2,left] {Supply};
\path [line] ([xshift=1.3em]WSM.north) -- ++(0,0.67) -- ++ (0,0) -- ([xshift=1.3em]Gens.south) [dashed] node [transition,pos=0.2,right] {Payment};

\path [line, line width=1.1] ([yshift=0.8em]Gens.west) -- ++(-1.2,0) -- ++ (0,-4.8) -- ([yshift=-0.8em]Ut.west) node [transition,pos=0.0,left] {$\boldsymbol{\bar{q}}$};
\path [line, line width=1.1] ([yshift=0.8em]Ut.west) -- ++(-0.4,0) -- ++ (0,3.68) -- ([yshift=-0.8em]Gens.west) node [transition,pos=0.15,left] {$\boldsymbol{\bar{\lambda}, P}$};
\end{tikzpicture}
\vspace{0.04cm}
\caption{Energy Market Participants and their Interactions}
\label{fig:Market_Participants}
\end{figure}
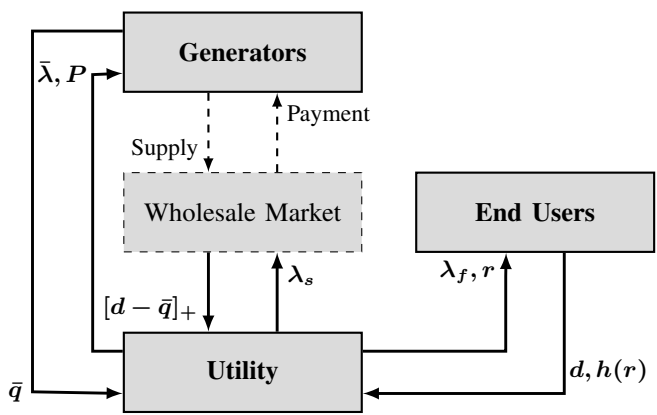

Figure \ref{fig:Market_Participants} illustrates the interaction between generating companies, load-serving entities (utilities), the wholesale electricity market, and end-users of electricity.

The electric utility can strike one-to-one contracts with generating companies to purchase a fixed amount of electricity $\bar{q}$ at a locked-in price $\bar{\lambda}$ to be delivered at some a-priori specified time in the future. $P$ denotes the premium for each reserved unit of electricity. The utility provides end-users with electricity at a fixed unit rate $\lambda_f$ and is obligated to cover the random demand $d$ at all times. The rate $\lambda_f$ is exogenously set by the Public Utilities Commission. However, the utility can use DR to incentivize users to temporarily reduce their demand. This is achieved by offering the reward $r$ to end-users, which elicits a demand reduction $h(r)$. If the demand $d$ exceeds $\bar{q}$, i.e. the purchased amount of electricity through one-to-one contracts with generators, the utility has to procure $[d-\bar{q}]_+$ units of electricity from the wholesale market at uncertain wholesale price $\lambda_s$ per unit. The market clearing price $\lambda_s$, reflected by Locational Marginal Prices (LMPs), is a random variable and depends on the ratio of energy supply by generators, the total demand $[d-\bar{q}]_+$, operational constraints, as well as congestion of the grid.

The interactions between generators and the utility as well as between end-users and the utility are instruments to hedge utilities against high prices $\lambda_s$. If the utility expects high wholesale prices $\lambda_s$, then it has an incentive to reduce customer demand $d$ by engaging in DR, or to procure cheaper electricity through contracts with generating companies. We make the following assumptions:

\begin{assumption}\label{as:risk_neutral}
The utility is risk-neutral.
\end{assumption}

\begin{assumption}\label{as:price_taking}
The utility is price-taking.
\end{assumption}
Assumption \ref{as:price_taking} is a natural assumption, stating that the utility cannot influence prices by exerting market power. Together with Assumption \ref{as:risk_neutral}, the question we seek to answer in the remainder of this paper is how the utility maximizes its expected profit in the presence of the random variables $d$ and $\lambda_s$ and hedging instruments. For simplicity, we focus on a single load zone to avoid spatial heterogeneity of LMPs.

%
%


\section{Optimal Hedging Strategies}
\label{sec:Options}

\begin{figure}[h]
\vspace*{-0.1cm}
\centering
\resizebox{\linewidth}{!}{

\begin{tikzpicture}
\draw[line width=0.25mm,->] (0,0) -- (11,0);
\foreach \x in {4.35, 14.4}{
   \draw [line width=0.25mm](\x/1.7,3pt) -- (\x/1.7,-3pt);
}
\draw (2.57,0) node[below=3pt] {\begin{small}
\begin{tabular}{c}
	$\bar{\lambda}, P, h(r)$ are announced.\\
    Utility purchases hedging\\
    instruments to maximize $\mathbb{E}[\Pi]$\\
    as function of random $d, \lambda_s$.
\end{tabular}
\end{small}} node[above=3pt] {{\small $t=0$}};
\draw (8.48,0) node[below=3pt] {\begin{small}
\begin{tabular}{c}
    $d$ and $\lambda_s$ materialize.\\
    Profit $\Pi$ is determined.\\
    Settlements between generators,\\
    LSE, and end users take place.
\end{tabular}
\end{small}} node[above=3pt] {{\small $t=1$}};
\end{tikzpicture}
}
\caption{Timeline of Hedging}
\label{fig:timeline}
\end{figure}
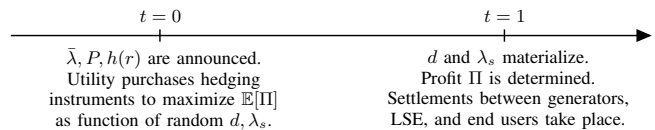

Let $\lambda_s$ and $d$ be random variables with cumulative distribution functions (CDF) $G(\cdot)$ and $F(\cdot)$, respectively. $G$ and $F$ are assumed to have support $[0, \infty)$ and $[d_{\min}, d_{\max}]$, respectively, where $0\leq d_{\min}\leq d_{\max}$. We assume the absence of any energy storage capabilities and focus on a single-period setting, where the LSE can purchase hedging instruments at time 0, possessing only an estimate of consumer demand $d$ and real-time spot price $\lambda_s$ at time 1. At time 1, the random variables $d$ and $\lambda_s$ materialize and the LSE's profit $\Pi$ as a function of the hedging instruments purchased at time 0 is determined. Figure \ref{fig:timeline} illustrates the hedging process. The LSE aims to maximize its \textit{expected} profit $\mathbb{E}[\Pi]$ by deciding on its portfolio of hedging instruments at $t=0$. 

In the following, we analyze the cases for (a) no hedging instruments, (b) forward contract, (c) call option, and (d) DR and derive explicit expressions for the optimal contracts and corresponding profits for cases (b)-(d). 

\subsection{Base Case (No Hedging)}
\label{sec:base_case}
If the LSE does not buy any options at stage 0, its expected profit at time 1 is simply
\begin{align}\label{eq:expected_profit_no_options}
\mathbb{E}[\Pi] = (\lambda_f - \mathbb{E}[\lambda_s])\cdot\mathbb{E}[d].
\end{align}
We will compare the profit of this base case to the forward contract, call option, and DR in the following.

\subsection{Forward Contract}
A forward contract is a one-on-one agreement between the LSE and an electricity generator, which obligates the generator (at time 0) to deliver a fixed amount of electricity $\bar{q}$ at a locked-in price $\bar{\lambda}_F$ to the LSE at some point in the future (time 1). Forward contracts possess high flexibility and are traded as over-the-counter products. The LSE seeks to sign such a contract if it has reason to believe the expected wholesale price at the time of delivery to exceed $\bar{\lambda}_F$, and the generator will do so in the opposite case. If, at time 1, $\bar{q} > d$, the LSE has purchased too much volume at time 0, and so $\bar{q}-d$ are wasted. Conversely, if $\bar{q} < d$ at time 1, $d-\bar{q}$ units of electricity have to be bought at real-time spot price $\lambda_s$.

The profit $\Pi_F$ under a forward contract of volume $\bar{q}$ at unit price $\bar{\lambda}$ is therefore expressed as
\begin{align}\label{eq:LSE_fixed_forward_contract}
\Pi_F = \lambda_f d - \bar{\lambda}_F \bar{q} - \lambda_s [d-\bar{q}]_+.
\end{align}

\begin{theorem}[Optimal Forward Contract]\label{thm:optimum_forward_contract}
With $\mathbb{E}[\lambda_s] > \bar{\lambda}_F$, the optimal contract volume $\bar{q}^\ast$ and the optimal expected profit $\mathbb{E}[\Pi_F^\ast]$ become
\begin{subequations}
\begin{align}
\bar{q}^\ast &= F^{-1}\left( 1-\frac{\bar{\lambda}_F}{\mathbb{E}[\lambda_s]} \right),\label{eq:opt_vol_forward_contract}\\
\mathbb{E}[\Pi_F^\ast] &= \lambda_f \mathbb{E}[d] - \mathbb{E}[\lambda_s] \int_{F^{-1}\left( 1 - \frac{\bar{\lambda}_F}{\mathbb{E}[\lambda_s]} \right)}^\infty xf(x)~dx.\label{eq:opt_profit_forward_contract}
\end{align}
\end{subequations}
\end{theorem}

\subsection{Call Option}
Similar to fixed forward contracts, the LSE can strike one-on-one deals with a counterparty over an agreed volume $\bar{q}$ at strike price $\bar{\lambda}_C$. The key difference is that the LSE can, but is not obliged to, exercise the call option if $\bar{\lambda}_C < \lambda_s$ at time 1. Typically the buyer of the call option pays a premium $P$ for each unit of the call option.

The profit $\Pi_C$ under a call option with volume $\bar{q}$ at strike price $\bar{\lambda}_C$ at the premium $P$ per unit can thus be written as
\begin{equation}
\begin{aligned}\label{eq:LSE_Call_Option_Return}
\Pi_C = &~\lambda_f d  - \lambda_s\left[d - \bar{q}\right]_+ - P\bar{q}\\
&- \min(\bar{\lambda}_C, \lambda_s)\cdot\min(d, \bar{q}).
\end{aligned}
\end{equation}
The last term of \eqref{eq:LSE_Call_Option_Return} encodes the fact that the LSE can cover up to $\bar{q}$ units at the cheaper of the strike price $\bar{\lambda}_C$ or the wholesale price $\lambda_s$. The remainder $[d-\bar{q}]_+$ has to be purchased from the spot market at price $\lambda_s$.

\begin{theorem}[Optimal Call]\label{thm:optimum_call}
With $\mathbb{E}[\lambda_s] > P+\bar{\lambda}_C - \int_0^{\bar{\lambda}_C}G(y)dy$, the profit-maximizing call volume $\bar{q}^\ast$ and the corresponding optimal expected profit $\mathbb{E}[\Pi_C^\ast]$ are
\begin{subequations}
\begin{align}
\bar{q}^\ast =&~ F^{-1}\left( 1 - \frac{P}{\mathbb{E}[\lambda_S] - \bar{\lambda}_C + \int_0^{\bar{\lambda}_C}G(y)dy} \right),\label{eq:opt_vol_call}\\
\mathbb{E}[\Pi_C^\ast] =&~ \left(\lambda_f - \bar{\lambda} + \int_0^{\bar{\lambda}_C}G(y)dy\right) \mathbb{E}[d]\label{eq:opt_profit_call}\\
&~ - \left(\mathbb{E}[\lambda_s] - \bar{\lambda}_C + \int_0^{\bar{\lambda}_C}G(y)dy\right) \int_{\bar{q}^\ast}^\infty x f(x)dx.\nonumber
\end{align}
\end{subequations}

\end{theorem}

\subsection{Demand Response}
We model the effect of demand response as a shift in the distribution of the consumer towards zero, induced by the monetary reward $r\in\mathbb{R}_+$ transferred from the LSE to the consumer as a lump sum. Note that the real reduction of the consumer in response to the DR signal has to be estimated by constructing the counterfactual consumption in the absence of the DR signal, whose estimation is beyond the scope of this paper. The interested reader is referred to \cite{Zhou:2016aa, Zhou:2016ab}.

Let $f(d)$ denote the probability density function of $d$ in the absence of any reward with support $[d_{\min}, d_{\max}]$. Let $F(d|r)$ denote the cumulative distribution function of the random variable $d$, given the reward level $r$. Then the distribution shift is modeled as
\begin{equation}\label{eq:distribution_shift}
F(d|r) = \begin{cases}
      0, & \mathrm{if}\ d < d_{\min}  \\
      F(d + h(r)), & \mathrm{if}\ d \geq d_{\min}
    \end{cases}
\end{equation}
where $h(r)$ is a concave, increasing function representing the elasticity of the user in response to reward $r$, i.e. the relative reduction of consumption as a function of $r$. $h(r)$ is equivalent to the shift of the location parameter of distribution $f(\cdot)$. We make the following
\begin{assumption}\label{as:linear_shift}
The reward $r \geq 0$ induces a linear shift, i.e.
\begin{equation}
h(r) = \alpha r,\quad \alpha > 0.
\end{equation}
\end{assumption}
With Assumption \ref{as:linear_shift} and the definition of the distribution shift, it becomes clear that the distribution $f(\cdot|r)$, given a reward $r>0$, has support $[d_{\min}, d_{\max}-h(r)]$ with discrete mass $\int_{d_{\max}-h(r)}^{d_{\max}}f(x)~dx$ at $d_{\min}$.

Assumption \ref{as:linear_shift} is necessary for analytical tractability of the DR hedging case. We note that the linearity of $h(r)$ is unrealistic, since it implies that for large enough reward levels $r$, the user consumes zero with probability 1. However, for small reward levels, a linear price elasticity of demand $h(r)$ can be justified.

The LSE's profit $\Pi_{\mathrm{DR}}$ with Demand Response is
\begin{equation}\label{eq:LSE_DR_Return}
\Pi_{\text{DR}} = (\lambda_f - \lambda_s) d(r) - r.
\end{equation}
From \eqref{eq:LSE_DR_Return}, it immediately follows that DR only makes sense in the presence of large expected spot prices $\mathbb{E}[\lambda_s]$ at time 1 which exceed the fixed contractual price $\lambda_f$. Then the optimal profit $\Pi_{\mathrm{DR}}^\ast$ is the minimal expected loss of the LSE.

\begin{theorem}[Optimal Demand Response]\label{thm:optimum_DR}
With $\mathbb{E}[\lambda_s] > \lambda_f$, the profit-maximizing reward $r^\ast$ and the optimal expected profit $\mathbb{E}\Pi_{\mathrm{DR}}^\ast$ are
\begin{subequations}
\begin{align}
r^\ast &= \begin{cases}
      \frac{1}{\alpha}F^{-1}\left( 1 - \frac{1}{\alpha\cdot \left(\mathbb{E}\left[ \lambda_s\right] - \lambda_f\right)} \right), & \mathrm{if}\ \frac{1}{\alpha} < \mathbb{E}[ \lambda_s] - \lambda_f  \\
      0, & \mathrm{otherwise}\label{eq:opt_reward_DR}
    \end{cases}\\
\mathbb{E}\Pi_{\mathrm{DR}}^\ast &= \begin{cases} (\lambda_f - \mathbb{E}[\lambda_s]) \int_{\alpha r^\ast}^\infty x f(x)dx, & \mathrm{if}\ \frac{1}{\alpha} < \mathbb{E}[ \lambda_s] - \lambda_f  \\
(\lambda_f - \mathbb{E}[\lambda_s])~\mathbb{E}[d], & \mathrm{otherwise}\label{eq:opt_profit_DR}
\end{cases}
\end{align}
\end{subequations}
\end{theorem}
The condition $\alpha > (\mathbb{E}[ \lambda_s] - \lambda_f)^{-1}$ for the optimal reward means that the ability to shift, $1/\alpha$, must be greater than the inverse of the expected price difference $(\mathbb{E}[ \lambda_s] - \lambda_f)^{-1}$ to make DR profitable. The higher the expected price difference $\mathbb{E}[ \lambda_s] - \lambda_f$, the less stringent the requirement on $\alpha$, which agrees with intuition.

\begin{theorem}[Diversified Portfolios]\label{thm:unique_portfolio}
For general demand distributions, the optimal portfolio can either consist of a unique option or a combination of call and forward contract options, but never of a combination of DR and either call or forward contract options. For the special case of a uniform demand distribution, the optimal portfolio always consists of a unique option, i.e. diversified portfolios consisting of more than one option are always suboptimal.
\end{theorem}
Depending on the properties of the demand distribution $F(\cdot)$, a mixed portfolio of call and forward contract options can exist, but is impossible to obtain in closed form for general distributions. This is consistent with the approach in \cite{Oum:2006aa} where the authors replicate the optimal portfolio (which would be continuous) with a finite set of options. Due to Theorem \ref{thm:unique_portfolio}, we restrict our attention to optimal portfolios consisting of a unique option in the remainder of this paper.

%
%

\section{The Effect of Uncertainty}
\label{sec:Uncertainty}
For a better understanding of the optimal profits under the different contracts $\Pi_{F}^\ast, \Pi_{C}^\ast, \Pi_{\mathrm{DR}}^\ast$ introduced in the previous section, we relate these quantities to properties of the consumption distribution $F(\cdot)$.

\subsection{Influence of Distribution Tail}
By incorporating the \textit{Conditional Value-at-Risk} (CVaR) measure \cite{Rockafellar:2002aa}, we can relate the optimal profits to the tail properties of the consumption density $f(\cdot)$. The CVaR at confidence level $\alpha\in(0,1)$ of a random variable $X$ with CDF $F(\cdot)$ representing loss is formally defined as
\begin{align}\label{eq:conditional_value_at_risk}
\text{CVaR}_\alpha(X) = \mathbb{E}[X~|~X\geq F^{-1}(\alpha)]
\end{align}
and can be interpreted as the expected loss attained in the worst $(1-\alpha)\cdot 100\%$ of cases or the expectation of the $(1-\alpha)$ probability tail of $X$. With this definition, the optimal expected profits under the different options $\Pi_{F}^\ast, \Pi_{C}^\ast$, and $\Pi_{\mathrm{DR}}^\ast$ are reformulated in Proposition \ref{prop:reformulation_with_cvar}.

\begin{proposition}\label{prop:reformulation_with_cvar}
With $\alpha > (\mathbb{E}[ \lambda_s] - \lambda_f)^{-1}$ and the definition of CVaR, the optimal expected profits under the forward contract $\mathbb{E}[\Pi_F^\ast]$, the call option $\mathbb{E}[\Pi_C^\ast]$, and Demand Response $\mathbb{E}[\Pi_{\mathrm{DR}}^\ast]$ can be expressed as follows:
\begin{subequations}
\begin{align}
\mathbb{E}[\Pi_F^\ast] =&~ \lambda_f \mathbb{E}[d] - \bar{\lambda}_F\mathbb{E}[d~|~d\geq F^{-1}(1-\bar{\lambda}_F/\mathbb{E}[\lambda_s])]\nonumber\\
=&~ \lambda_f \mathbb{E}[d] - \bar{\lambda}_F\cdot\mathrm{CVaR}_{\alpha_F}(d)\label{eq:opt_profit_forward_cvar}\\
\mathbb{E}[\Pi_C^\ast] =&~ \left(\lambda_f - \bar{\lambda}_C + \int_0^{\bar{\lambda}_C}G(y)dy\right)\mathbb{E}[d]\label{eq:opt_profit_call_cvar}\\
&~- P\cdot\mathrm{CVaR}_{\alpha_C}(d)\nonumber\\
\mathbb{E}[\Pi_{\mathrm{DR}}^\ast] =&~  - \frac{1}{\alpha}\cdot\mathrm{CVaR}_{\alpha_{\mathrm{DR}}}(d)\label{eq:opt_profit_dr_cvar}
\end{align}
\end{subequations}
where we used the definitions
\begin{subequations}
\begin{align}
\alpha_{F} &= 1 - \frac{\bar{\lambda}_F}{\mathbb{E}[\lambda_s]}\label{eq:alpha_forward_contract}\\
\alpha_{C} &= 1 - \frac{P}{\mathbb{E}[\lambda_s] - \bar{\lambda}_C + \int_0^{\bar{\lambda}_C}G(y)dy}\label{eq:alpha_call_option}\\
\alpha_{\mathrm{DR}} &= 1 - \frac{1}{\alpha\cdot \left(\mathbb{E}[ \lambda_s] - \lambda_f\right)}\label{eq:alpha_demand_response}
\end{align}
\end{subequations}
\end{proposition}
From Proposition \ref{prop:reformulation_with_cvar}, it follows that the optimal profit decreases as the conditional expectation of the tail increases, that is, the more heavy-tailed the consumption distribution $f(\cdot)$ becomes. It is illustrative to analyze the optimal decisions and corresponding optimal expected profits for perfect information of $d$, which are given in the following:
\begin{subequations}
\begin{align}
\bar{q}_F^\ast|d =&~ d, \quad\quad \bar{q}_C^\ast|d = d,\quad\quad r^\ast|d = d/\alpha\nonumber\\
\mathbb{E}[\Pi_F^\ast|d] =&~ (\lambda_f-\bar{\lambda}_F)\cdot d\label{eq:opt_profit_forward_perfect}\\
\mathbb{E}[\Pi_C^\ast|d] =&~ \left(\lambda_f - \bar{\lambda}_C + \int_0^{\bar{\lambda}_C}G(y)dy-P\right)d\label{eq:opt_profit_call_perfect}\\
\mathbb{E}[\Pi_{\text{DR}}|d] =&~ -d/\alpha\label{eq:opt_profit_DR_perfect}
\end{align}
\end{subequations}
$\bar{q}_F^\ast|d$ and $\bar{q}_C^\ast|d$ denote the optimal forward contract and call volume, respectively. $r^\ast|d$ signifies the optimal DR reward.

\subsection{Influence of Statistical Dispersion}
In this section, we attempt to construct a relationship between the statistical dispersion of the consumption distribution $F(\cdot)$ and the optimal expected profit. Intuitively, the more spread out the distribution $F(\cdot)$, the lower the expected profit. While many measures for statistical dispersion exist in the literature, such as interquartile ranges, absolute deviation, variance-to-mean-ratio, etc., we express the optimal expected profits $\mathbb{E}[\Pi_F^\ast]$, $\mathbb{E}[\Pi_C^\ast]$, and $\mathbb{E}[\Pi_{\mathrm{DR}}^\ast]$ in terms of the standard deviation $\sigma$ for the special case of a uniform distribution with support $[d_{\min}, d_{\max}]$ for expositional ease and analytical tractability.

\begin{proposition}\label{prop:reformulation_with_std}
For the uniform distribution $F(\cdot)$ with support $[d_{\min}, d_{\max}]$, the optimal expected profits under the conditions $\mathbb{E}[\lambda_s] > \max\left(\bar{\lambda}_F, P+\bar{\lambda}_C-\int_{0}^{\bar{\lambda}_C} G(y)~dy\right)$ and $\alpha > (\mathbb{E}[\lambda_s]-\lambda_f)^{-1}$ are expressed as follows:
\begin{subequations}
\begin{align}
\mathbb{E}[\Pi_F^\ast] &= \lambda_f \mathbb{E}[d] - \bar{\lambda}_F d_{\min} - \sqrt{3}\mathbb{E}[\lambda_s](1-\alpha_F^2)\sigma \label{eq:profit_dispersion_forward}\\
\mathbb{E}\Pi_C^\ast &= \left(\lambda_f - \bar{\lambda}_C + \int_0^{\bar{\lambda}_C}\hspace{-0.27cm}G(y)dy\right)\mathbb{E}[d] - Pd_{\min}\label{eq:profit_dispersion_call}\\
&-\sqrt{3}\left(\mathbb{E}[\lambda_s] - \bar{\lambda}_C + \int_0^{\bar{\lambda}_C}G(y)dy\right)(1-\alpha_{C}^2)\sigma\nonumber\\
\mathbb{E}[\Pi_{\mathrm{DR}}^\ast] &= -d_{\min}/\alpha - \sqrt{3}(\mathbb{E}[\lambda_s]-\lambda_f)(1-\alpha_{\mathrm{DR}}^2)\sigma\label{eq:profit_dispersion_demand_response}
\end{align}
\end{subequations}
\end{proposition}

For the case of perfect information, i.e. $\sigma=0$ and $d_{\min}=d_{\max}=d$, the equations for the optimal expected profit under perfect information \eqref{eq:opt_profit_forward_perfect}-\eqref{eq:opt_profit_DR_perfect} are recovered. Equations \eqref{eq:profit_dispersion_forward}-\eqref{eq:profit_dispersion_demand_response} explain that the optimal expected profit for each case decreases linearly in $\sigma$, giving rise to the notion that more ``spread out'' distributions diminish the expected profit. The rate of decrease depends on case-specific parameters, whose relation to each other determines which hedging option is profit-maximizing for a particular case. As consumption distributions typically are plagued by a large amount of uncertainty (large $\sigma$), improved load predictions to decrease $\sigma$ have a direct economic benefit to the utility.

%
%

\section{Choosing the Best Option}
\label{sec:best_option}
We now derive conditions on the random variables $\lambda_s$ and $d$ with distributions $G(\cdot)$ and $F(\cdot)$ and the option parameters $\bar{\lambda}_F$, $\bar{\lambda}_C$, $P$, and $\alpha$ announced at time 0 to determine the best hedging strategy consisting of a unique option. For analytical tractability, we make the following assumptions:

\begin{assumption}\label{as:spot_price_uniform_distribution}
The real-time spot price $\lambda_s$ is uniformly distributed with support $[0, s_{\max}]$, i.e. $G(y) = \frac{1}{s_{\max}}\mathbf{1}_{0\leq y\leq s_{\max}}$.
\end{assumption}

\begin{assumption}\label{as:consumption_uniform_distribution}
The consumption is uniformly distributed in $[0, d_{\max}]$, i.e. $F(x) = \frac{1}{d_{\max}}\mathbf{1}_{0\leq x\leq d_{\max}}$.
\end{assumption}

\begin{theorem}\label{thm:best_option_choice}
Under Assumptions \ref{as:spot_price_uniform_distribution} and \ref{as:consumption_uniform_distribution} and $\mathbb{E}[\lambda_s] > \lambda_f$, the forward contract is preferred over the call option, if
\begin{align}\label{eq:forward_call_ideal}
\bar{\lambda}_F \leq \mathbb{E}[\lambda_s] - \frac{\mathbb{E}[\lambda_s]-\bar{\lambda}_C + \bar{\lambda}_C^2/(4\mathbb{E}[\lambda_s])-P}{\sqrt{1-\frac{\bar{\lambda}_C - \bar{\lambda}_C^2/(4\mathbb{E}[\lambda_s])}{\mathbb{E}[\lambda_s]}}}.
\end{align}
DR is preferred over the forward contract, if
\begin{align}\label{eq:forward_DR_ideal}
\frac{1}{\alpha} \leq (\mathbb{E}[\lambda_s]-\lambda_f)\left[ 1 - \sqrt{\frac{\mathbb{E}[\lambda_s]}{\mathbb{E}[\lambda_s]-\lambda_f}}\left( 1-\frac{\bar{\lambda}_F}{\mathbb{E}[\lambda_s]} \right) \right].
\end{align}
Finally, DR is preferred over the call option, if
\begin{align}\label{eq:call_DR_ideal}
\frac{1}{\alpha} \leq (\mathbb{E}[\lambda_s]-\lambda_f)\left[ 1-\sqrt{\frac{L}{(\mathbb{E}[\lambda_s]-\lambda_f)}}\left( 1-\frac{P}{L} \right) \right].
\end{align}
with $L = (\mathbb{E}[\lambda_s]-\bar{\lambda}_C+\bar{\lambda}_C^2)/(4\mathbb{E}[\lambda_s])$ and where $\bar{\lambda}_F$ and $\bar{\lambda}_C$ denote the unit price for each reserved unit of electricity under the forward contract and the call option, respectively.
\end{theorem}

%
%

\section{Simulations}
\label{sec:simulations}
Assumptions \ref{as:spot_price_uniform_distribution} and \ref{as:consumption_uniform_distribution} admitted a closed form solution to the best hedging instrument, stated in \eqref{eq:forward_call_ideal}-\eqref{eq:call_DR_ideal}. For a more elaborate analysis, we now repeat this exercise by approximating the demand distribution $F(\cdot)$ as well as the distribution of spot prices $G(\cdot)$ with real data from California to approximate decision boundaries for which the expected profits under different hedging instruments are identical. Since closed-form solution under this more realistic scenario do not exist, we plot these optimal decision boundaries as a function of the hedging parameters $P, \bar{\lambda}_F, \bar{\lambda}_C$, and $\alpha$.

\subsection{Empirical Distribution of Demand}
\begin{figure}[hbtp]
\centering
\vspace*{-0.5cm}
\includegraphics[scale=0.29]{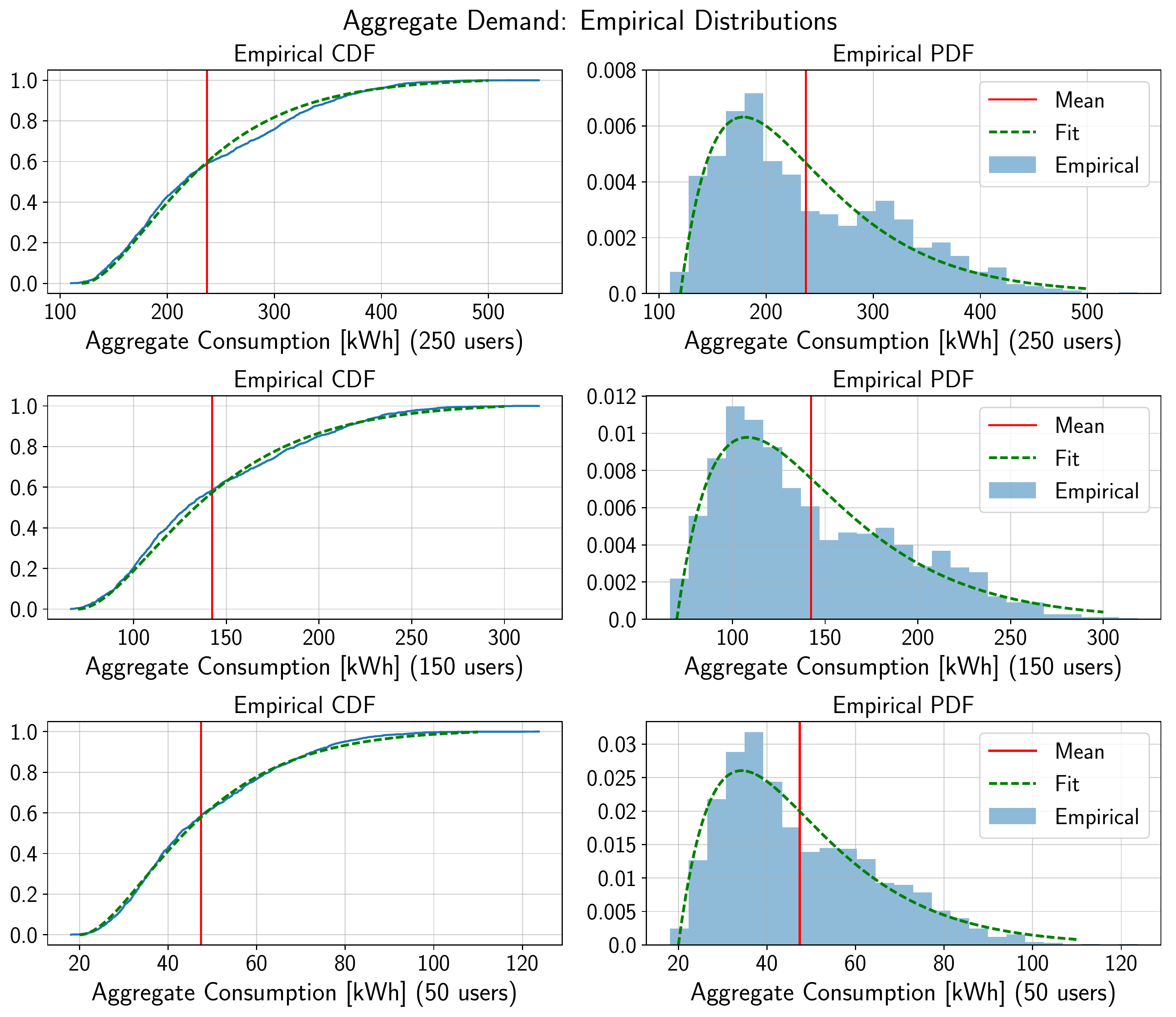}
\caption{Distribution of Aggregate Hourly Consumption for Varying Aggregation Sizes, 4-6 pm. Top: 250 Users, Middle: 150 Users, Bottom: 50 Users.}
\label{fig:Consumption_Distribution}
\vspace*{-0.5cm}
\end{figure}

We use hourly smart meter data from residential customers in California from the utilities Pacific Gas \& Electric, San Diego Gas \& Electric, and Southern California Edison to create a demand distribution for different sizes of user aggregations. The observations are restricted to hourly consumptions between 4-5 pm and 5-6 pm. Figure \ref{fig:Consumption_Distribution} shows the empirical PDFs and CDFs for different sizes of user aggregations. We approximate both functions as follows:
\begin{subequations}
\begin{align}
\hat{f}(x) &= a(x-d_{\min})e^{-cx},\quad a, c \in\mathbb{R}_+, x\in[d_{\min}, d_{\max}] \label{eq:approx_density_function_consumption}\\
\hat{F}(x) &= \frac{a}{c^2}\left(cd_{\min}-cx-1\right)e^{-cx} + \gamma, \quad \gamma\in\mathbb{R}\label{eq:approx_distribution_function_consumption}
\end{align}
\end{subequations}
With the constraints $\hat{F}(d_{\min}) = 0$ and $\hat{F}(d_{\max}) = 1$, the parameters $a$ and $\gamma$ can be found as a function of the decay parameter $c$. 
It can be seen that the approximations \eqref{eq:approx_density_function_consumption} and \eqref{eq:approx_distribution_function_consumption} fit the observed data reasonably well.

\subsection{Empirical Distribution of Wholesale Prices}
\begin{figure}[hbtp]
\centering
\includegraphics[scale=0.29]{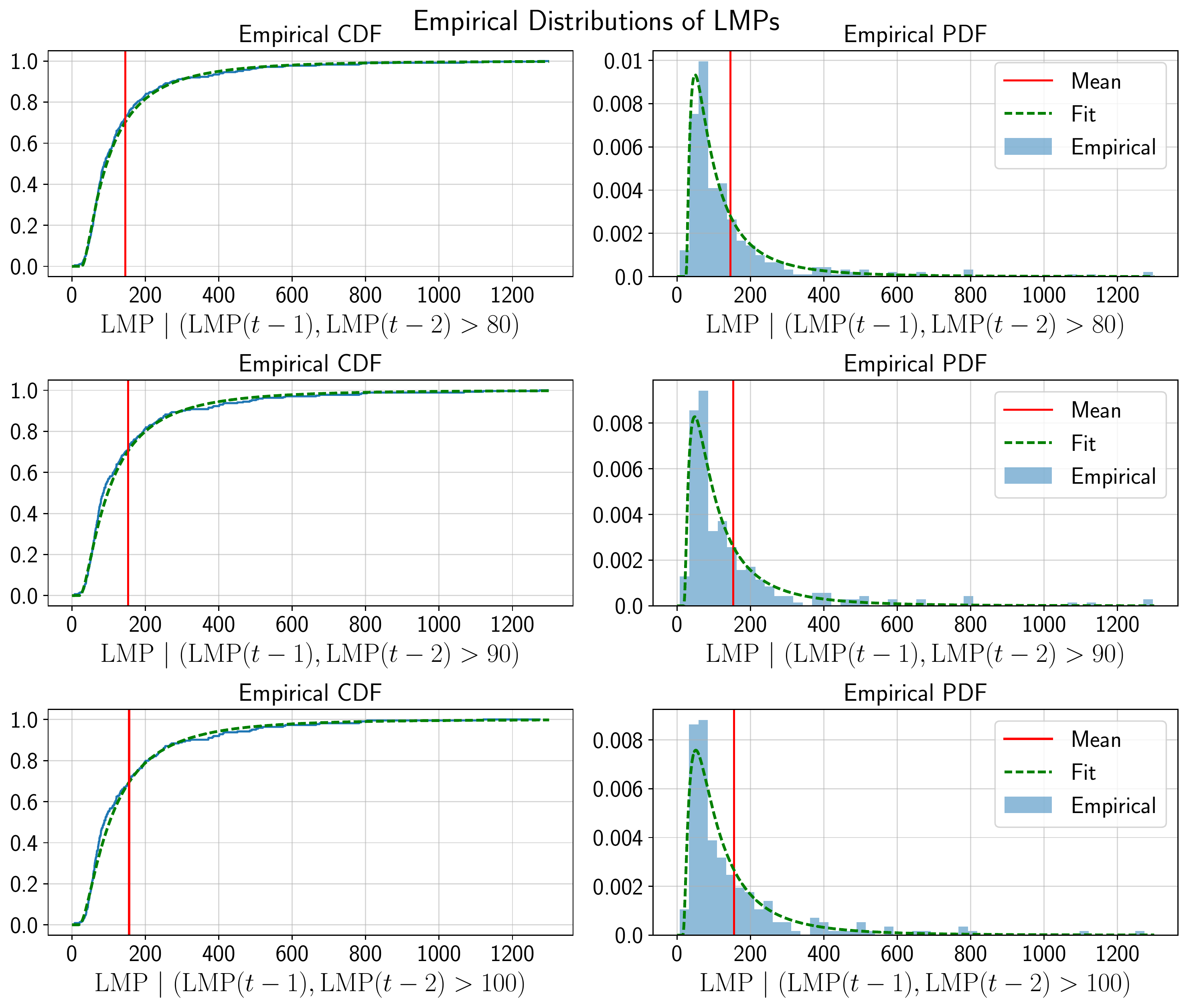}
\caption{Distributions of CAISO LMPs conditional on previous prices exceeding threshold $\xi$ for $\xi \in \lbrace 80\frac{\text{USD}}{\text{MWh}}, 90\frac{\text{USD}}{\text{MWh}}, 100\frac{\text{USD}}{\text{MWh}}\rbrace$.}
\label{fig:CAISO_LMPs_distribution}
\vspace*{-0.2cm}
\end{figure}

To obtain the price distribution $G(\cdot)$, we convert 5-minute locational marginal prices (LMPs) $\lambda_s$ set by the California Independent System Operator into an hourly format. The distribution $G(\cdot)$ of ``high'' LMPs is obtained by fitting a density function to the normalized histogram of those LMPs for which the two previous LMPs exceed the threshold $\xi > 0$, i.e. we consider all $\lbrace \lambda_s | \lambda_{s,t-1} \geq \xi, \lambda_{s,t-2} \geq \xi \rbrace$ for different thresholds $\xi$. We approximate the density function with a log-normal distribution:
\begin{align}\label{eq:lognormal_distribution}
\mathcal{N}(\ln x; \mu,\sigma) = \frac{1}{\sigma\sqrt{2\pi}}\exp\left( -\frac{(\ln x-\mu)^2}{2\sigma^2} \right)
\end{align}
which has support $[0, \infty)$, that is, we disregard negative LMPs. Figure \ref{fig:CAISO_LMPs_distribution} shows the observed data and the approximations for thresholds $\xi=80, 90, 100\frac{\text{USD}}{\text{MWh}}$.

\subsection{Pairwise Comparison of Hedging Instruments}
We now compute decision boundaries of equal expected profit for all 3 pairs of hedging instruments with Newton's method, using the demand and price distributions derived in \eqref{eq:approx_density_function_consumption}, \eqref{eq:approx_distribution_function_consumption}, and \eqref{eq:lognormal_distribution}.
\subsubsection{DR vs. Forward Contract}
\begin{figure}[hbtp]
\centering
\vspace*{0.1cm}
\includegraphics[scale=0.29]{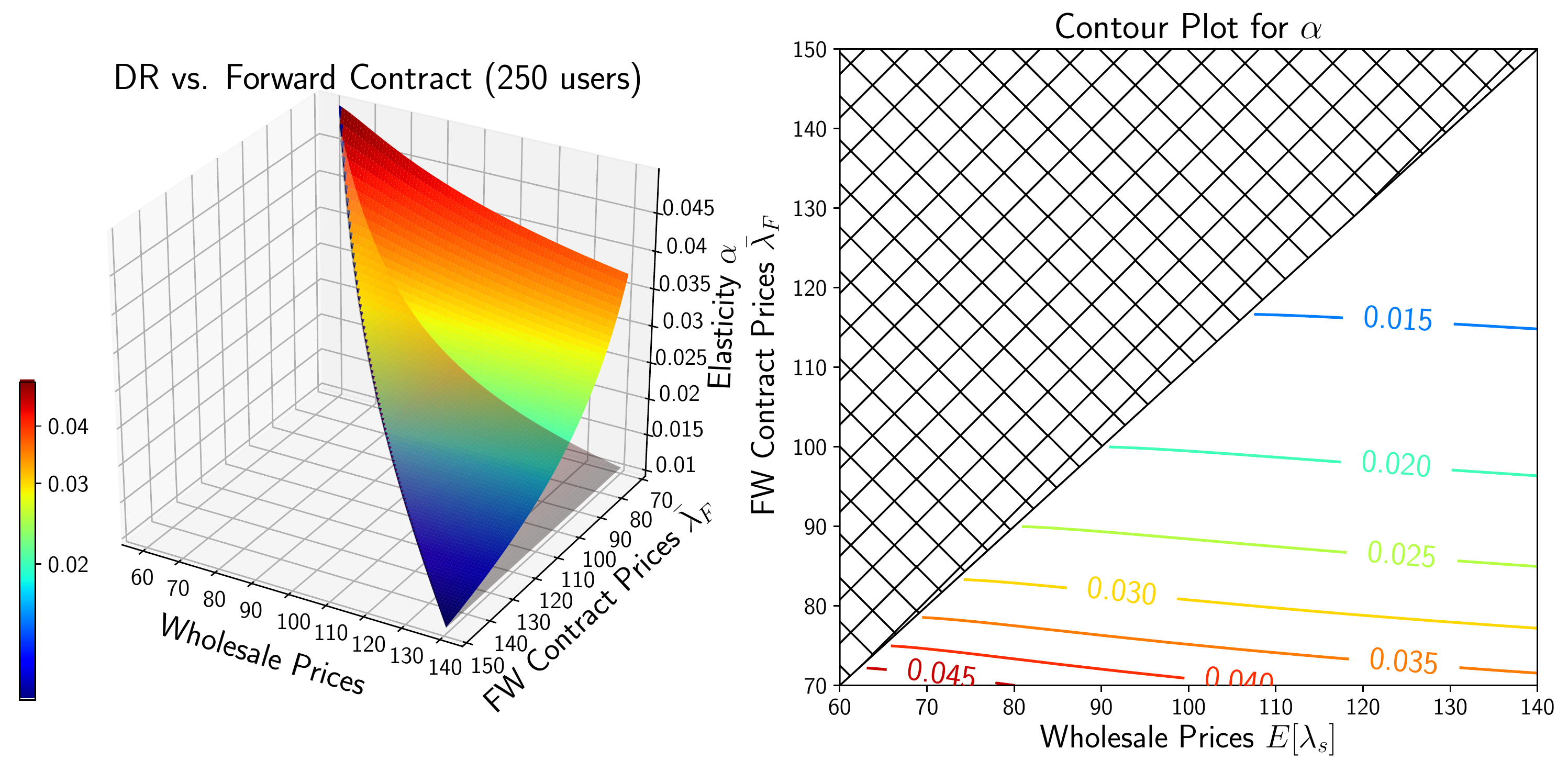}
\caption{Boundaries and contours of equal expected profit for forward option and DR, 250 users, $\lambda_f=0.05\frac{\text{USD}}{\text{kWh}}$.}
\label{fig:DR_vs_FW}
\vspace*{-0.4cm}
\end{figure}

Figure \ref{fig:DR_vs_FW} shows the decision boundary of elasticity $\alpha$ above which the optimal expected profits under DR is greater than under the forward contract, that is, $\mathbb{E}[\Pi_{\text{DR}}] \geq \mathbb{E}[\Pi_{F}]$, for different expected spot prices $\mathbb{E}[\lambda_s]$ and forward contract prices $\bar{\lambda}_F$, assuming $\lambda_f \leq \mathbb{E}[\lambda_s]$.
It is observed that $\alpha$ decreases as $\bar{\lambda}_F$ or the expected wholesale price $\mathbb{E}[\lambda_s]$ increase. The negative correlation of $\alpha$ with $\bar{\lambda}_F$ is consistent with expectations as a higher $\bar{\lambda}_F$ makes forward contracts more expensive. The fact that decreasing wholesale prices $\mathbb{E}[\lambda_s]$ make DR more competitive than forward contracts can be explained by comparing \eqref{eq:opt_profit_forward_contract} to \eqref{eq:opt_profit_DR}, which states that the entire demand $d$ has to be covered at price $\lambda_s$ in the DR case, compared to only $[d-\bar{q}]_+$ in the forward contract case. Also shown in Figure \ref{fig:DR_vs_FW} is the lower bound on $\alpha$ (gray transparent surface) below which DR is non-profitable, i.e. $\lbrace (\mathbb{E}[\lambda_s]-\lambda_f)^{-1}~|~70 \leq \mathbb{E}[\lambda_s] \leq 150 \rbrace$, where we set the residential tariff to $\lambda_f = 0.05~\text{USD/kWh}$.

\subsubsection{DR vs. Call}
\begin{figure}[hbtp]
\centering
\vspace*{0.1cm}
\includegraphics[scale=0.29]{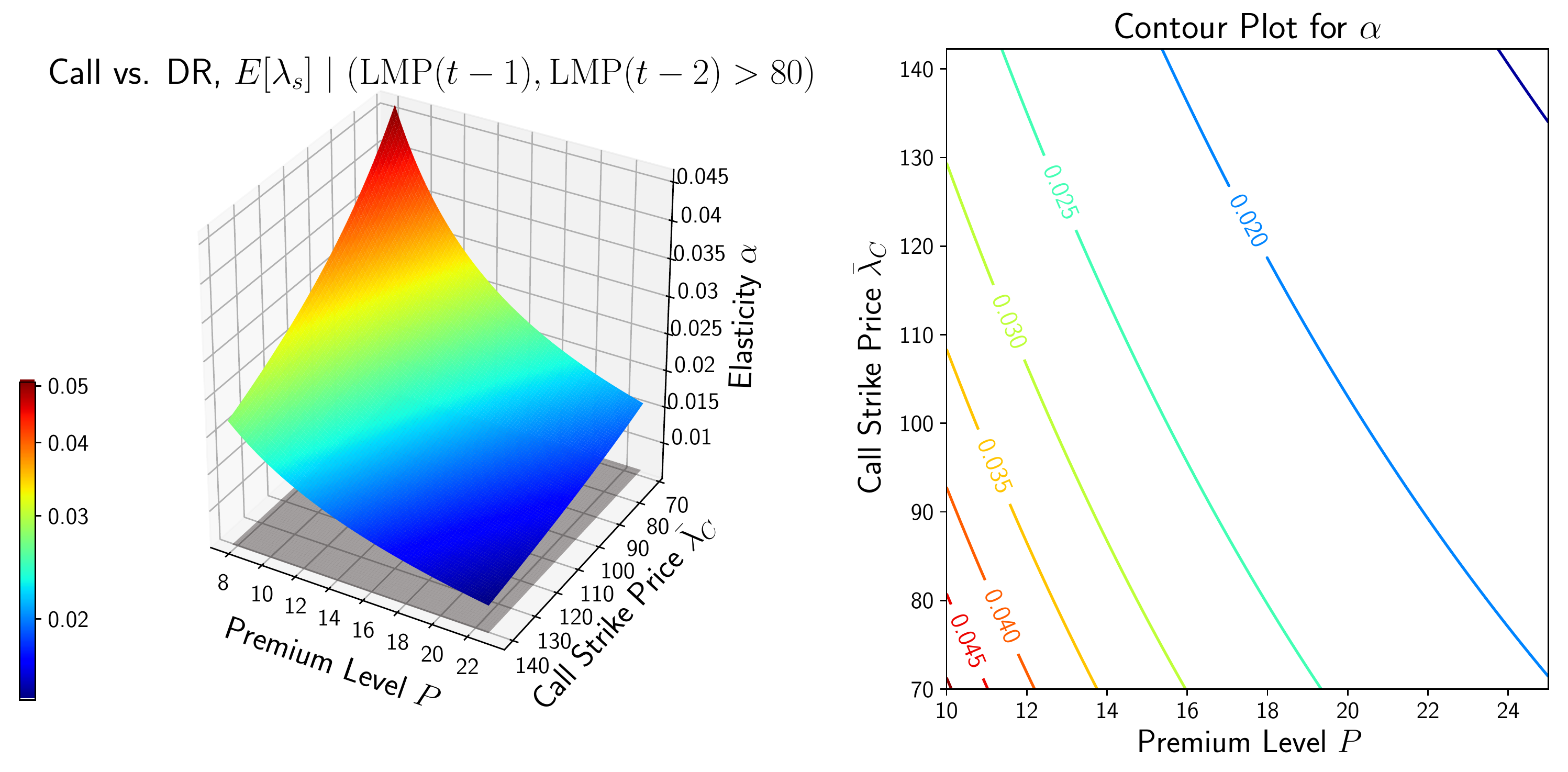}
\caption{Boundaries and contours of equal expected profit for DR and call option, 250 users, $\lambda_f=0.05\frac{\text{USD}}{\text{kWh}}$.}
\label{fig:DR_vs_Call}
\vspace*{-0.4cm}
\end{figure}

Figure \ref{fig:DR_vs_Call} shows the decision boundary of $\alpha$ for different call strike prices $\lambda_C$ and premium levels $P$ above which $\mathbb{E}[\Pi_{\text{DR}}] \geq \mathbb{E}[\Pi_{C}]$ with $\xi=80$. As the premium and strike price for the call option increase (and hence the call option becomes less attractive), DR becomes more profitable because $\alpha$ decreases.

\subsubsection{Forward Contract vs. Call}
\begin{figure}[hbtp]
\centering
\vspace*{-0.3cm}
\includegraphics[scale=0.29]{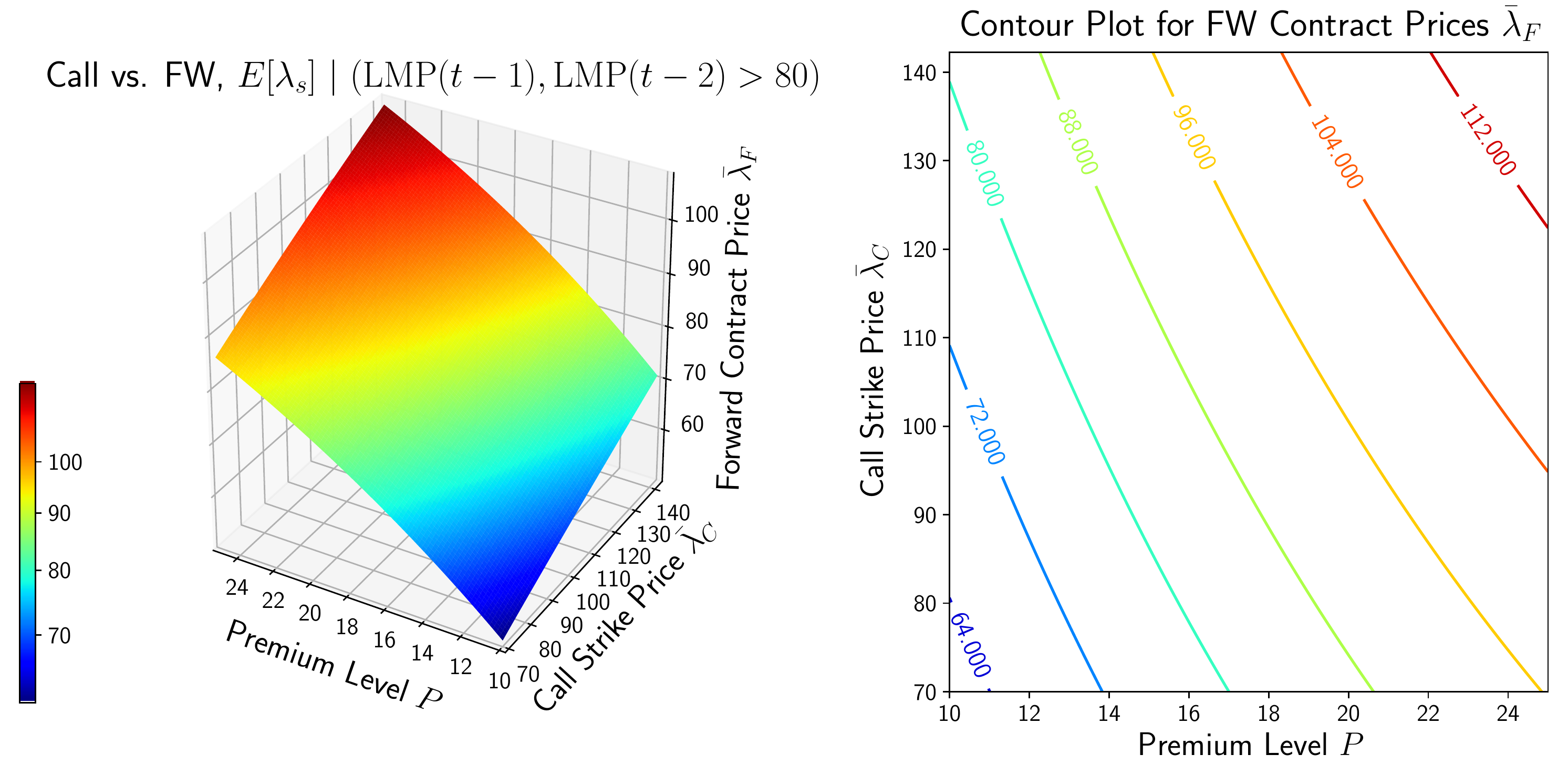}
\caption{Boundaries and contours of equal expected profit for forward and call option, 250 users, $\lambda_f=0.05\frac{\text{USD}}{\text{kWh}}$.}
\label{fig:FW_vs_call}
\vspace*{-0.3cm}
\end{figure}

Lastly, Figure \ref{fig:FW_vs_call} shows the decision surface for $\bar{\lambda}_F$ as a function of the call option parameters $P$ and $\bar{\lambda}_C$ above which the forward contract is more profitable in expectation, i.e. $\mathbb{E}[\Pi_{F}] \geq \mathbb{E}[\Pi_C]$. As expected, the forward contract becomes more attractive as either the premium $P$ or the call strike price $\bar{\lambda}_C$ increase.

\subsection{Evaluation}
Assuming a residential tariff of $0.05\frac{\text{USD}}{\text{kWh}}$, a lower bound on the elasticity $\alpha$ of approximately $0.02\frac{\text{MWh}}{\text{USD}} = 20\frac{\text{kWh}}{\text{USD}}$ at first glance seems to be an unachievable goal. However, note that wholesale prices can spike at up to $1000\frac{\text{USD}}{\text{MWh}}$, which is far outside the range of our calculations. Further, we disregarded transmission losses and capacity costs inherent to generators and utilities, which make the delivery of electricity under the forward contract and the call option more expensive, thereby lowering the bound on $\alpha$.

%
%


\section{Conclusion}
\label{sec:Conclusion}
We analyzed hedging instruments for load-serving entities to mitigate price risks associated with volatile energy supply and demand. Hedging against such risks is motivated by the fact that load-serving entities are obligated to meet energy demand of customers under contract instantaneously, which, in the absence of any hedging instruments, has to be procured in its entirety from the wholesale electricity market (at potentially high prices). Forward contracts and call options between load-serving entities and generating companies as well as Demand Response programs for end-users are methods to share this risk with other market participants. We formulated the optimal hedging strategy as a profit maximization problem which is random in the aggregate demand and wholesale electricity price. The optimal expected profit under each hedging instrument was found to be monotonely decreasing in the statistical dispersion of the demand distribution, and linearly decreasing for the special case of a uniform distribution. Using smart meter consumption data and locational marginal prices in California, we compared the optimal expected profits between the hedging methods in a pairwise fashion to generate decision boundaries of equal profit.

Our results can be extended in several regards. Firstly, a more involved analysis that takes into account operational constraints of the smart grid, e.g. transmission capacities and grid congestion, would add credibility to the suggestions of this paper. Secondly, analyzing how the optimal expected profit increases as a function of diminished uncertainty in electric wholesale prices and aggregate consumer demand due to forecasting is interesting from the perspective of profit maximization. Lastly, forgoing Assumptions \ref{as:risk_neutral} and \ref{as:price_taking} to allow utilities or generating companies to exercise market power calls for a game-theoretic formulation of the profit-maximization problem from the perspective of both generating companies and utilities, where each player seeks bids from the other in a mechanism design framework.


\bibliographystyle{IEEEtran}
\bibliography{bibliography}

%
%

\section*{Appendix}
\label{sec:Appendix}

\begin{lemma}[Leibniz Integral Rule]
For a function $f(x,t)$ with both $f(x,t)$ and $\frac{\partial f}{\partial x}$ continuous in $t\in [a(x), b(x)]$ and $x\in [x_0, x_1]$, where $a(x)$ and $b(x)$ are continuous in $x\in [x_0, x_1]$, for $x\in [x_0, x_1]$:
\begin{align*}
\frac{d}{dt}\left( \int_{a(t)}^{b(t)} f(x,t)~dx \right) =&
~ \int_{a(t)}^{b(t)} \frac{\partial f}{\partial t}dx + f(b(t),t)\cdot b^\prime(t)\\
& - f(a(t),t)\cdot a^\prime(t)
\end{align*}
\end{lemma}

\subsection*{Proof of Theorem \ref{thm:optimum_forward_contract}}
Taking the expectation of \eqref{eq:LSE_fixed_forward_contract} with respect to the random variables $\lambda_s$ and $d$ yields:
\begin{align}
\mathbb{E}[\Pi_F] =&~ -\bar{q}\cdot\bar{\lambda}_F + \lambda_f\int_0^{\bar{q}} xf(x)~dx + \lambda_f\bar{q}(1-F(\bar{q})) \nonumber\\
 &+ (\lambda_f - \mathbb{E}[\lambda_s])\int_{\bar{q}}^\infty (x-\bar{q})f(x)~dx \label{eq:expectation_forward_contract}
\end{align}
With the Leibniz Integral Rule, its derivatives with respect to $\bar{q}$ are
\begin{align*}
\frac{d\mathbb{E}[\Pi_F]}{d\bar{q}} &= -\bar{\lambda}_F + \lambda_f (1-F(\bar{q}) + (\lambda_f - \mathbb{E}[\lambda_s])(F(\bar{q})-1)\\
\frac{d^2\mathbb{E}[\Pi_F]}{d\bar{q}^2} &= -\lambda_f f(\bar{q}) + f(\bar{q})(\lambda_f - \mathbb{E}[\lambda_s]) < 0
\end{align*}
from which the optimal contract volume $\bar{q}^\ast$ \eqref{eq:opt_vol_forward_contract} follows. Plugging $\bar{q}^\ast$ back into \eqref{eq:expectation_forward_contract} yields
\begin{align*}
\mathbb{E}[\Pi_F] =& -\bar{\lambda}_F F^{-1}\left( 1 - \frac{\bar{\lambda}_F}{\mathbb{E}[\lambda_s]} \right) + \lambda_f \int_0^{\bar{q}}xf(x)~dx \\
&+ \frac{\lambda_f \bar{\lambda}_F}{\mathbb{E}[\lambda_s]}~F^{-1}\left( 1 - \frac{\bar{\lambda}_F}{\mathbb{E}[\lambda_s]} \right),
\end{align*}
from which the optimal profit \eqref{eq:opt_profit_forward_contract} follows.

\subsection*{Proof of Theorem \ref{thm:optimum_call}}
Similar to the previous proof, we take the expectation of \eqref{eq:LSE_Call_Option_Return} with respect to $\lambda_s$ and $d$:
\begin{align*}
&\mathbb{E}[\Pi_C] = \lambda_f\mathbb{E}[d] - \int_0^{\bar{q}}x f(x)dx \int_0^{\bar{\lambda}_C} y g(y)dy - P\bar{q} - r\\
&- \bar{\lambda}_C (1-G(\bar{\lambda}_C))\int_0^{\bar{q}}\hspace{-0.15cm} x f(x)dx  -  \bar{q}(1-F(\bar{q}))\int_0^{\bar{\lambda}_C}\hspace{-0.25cm} y g(y)dy\\
&- \bar{q}(1-F(\bar{q}))(1-G(\bar{\lambda}_C))\bar{\lambda}_C - \mathbb{E}[\lambda_S]\int_{\bar{q}}^\infty \hspace{-0.15cm} (x-\bar{q})f(x)dx
\end{align*}
The first order optimality condition reads
\begin{align*}
\frac{d\mathbb{E}\Pi_C}{d\bar{q}} =&~ -P + \mathbb{E}[\lambda_s](1-F(\bar{q})) \\
&~- (1-F(\bar{q}))\left[ \int_0^{\bar{\lambda}_C}\hspace{-0.25cm} y g(y) dy + \bar{\lambda}_C(1-G(\bar{\lambda}_C)) \right],
\end{align*}
which yields \eqref{eq:opt_vol_call} at the optimum. To show that this is a maximum, we compute the second derivative:
\begin{align*}
\frac{d^2\mathbb{E}\Pi_C}{d\bar{q}^2} =&~ f(\bar{q})\left[ \int_0^{\bar{\lambda}_C}\hspace{-0.25cm}y g(y) dy + \bar{\lambda}_C (1-G(\bar{\lambda}_C)) -\mathbb{E}[\lambda_s]\right],
\end{align*}
which is negative as we show below:
\begin{align*}
\int_0^{\bar{\lambda}_C}y g(y) dy + \bar{\lambda}_C(1-G(\bar{\lambda}_C)) &\stackrel{?}{<} \mathbb{E}[\lambda_s]\\
\bar{\lambda}_C G(\bar{\lambda}_C) - \int_0^{\bar{\lambda}_C}G(y)dy + \bar{\lambda}_C - \bar{\lambda}_C G(\bar{\lambda}_C) &\stackrel{?}{<} \mathbb{E}[\lambda_s]\\
0\leq \bar{\lambda}_C - \int_0^{\bar{\lambda}_C}G(y)dy <\bar{\lambda}_C &< \mathbb{E}[\lambda_s]
\end{align*}
Finally, the optimal expected profit $\mathbb{E}[\Pi_C^\ast]$ \eqref{eq:opt_profit_call} follows from plugging \eqref{eq:opt_vol_call} back into the expectation of \eqref{eq:LSE_Call_Option_Return}.

\subsection*{Proof of Theorem \ref{thm:optimum_DR}}
Taking the expectation of \eqref{eq:LSE_DR_Return} with respect to $\lambda_s$ and $r$ by performing Lebesgue-Stieltjes Integration gives
\begin{align}
\mathbb{E}[\Pi_{\text{DR}}] =&~ (\lambda_f - \mathbb{E}[\lambda_s])\int_{d_\text{min}}^{d_\text{max}-h(r)} xf(x+h(r))dx - r \nonumber\\
&~+ (\lambda_f - \mathbb{E}[\lambda_s])~d_{\text{min}}\int_{d_\text{max}-h(r)}^{d_\text{max}} f(x) dx \label{eq:expectation_DR_profit}\\
=&~ (\lambda_f - \mathbb{E}[\lambda_s])\int_{d_{\text{min}}+h(r)}^{d_\text{max}} (x-h(r))f(x)dx - r \nonumber
\end{align}
where we used the change of variables $x+h(r)\rightarrow x$ and the fact that $F(d_{\text{max}}) = F(d_{\text{max}}-h(r)) = 1$. With the Leibniz Integral Rule, its derivatives with respect to $r$ read
\begin{align*}
\frac{d\mathbb{E}[\Pi_{\text{DR}}]}{dr} &= (\lambda_f-\mathbb{E}[\lambda_s]) [1-F(h(r))](-h^\prime(r))-1\\
\frac{d^2\mathbb{E}[\Pi_{\text{DR}}]}{dr^2} &= (\underbrace{\lambda_f - \mathbb{E}[\lambda_s]}_{\leq 0})[ \underbrace{f(h)h^\prime + (F(h) - 1)h^{\prime\prime}}_{\geq 0} ]\Big|_{h=h(r)}
\end{align*}
For the linear shift, i.e. $h(r) = \alpha r$, first order optimality yields \eqref{eq:opt_reward_DR}, which is valid only under the condition that $\alpha > (\mathbb{E}[\lambda_s]-\lambda_f)^{-1}$. The second derivative is negative due to the concavity of $h(r)$, which results in $h^{\prime\prime}(r) \leq 0$. The optimal profit $\Pi_{\text{DR}}^\ast$ follows from plugging $r^\ast$ back into \eqref{eq:expectation_DR_profit}:
\begin{align*}
\mathbb{E}[\Pi_{\text{DR}}^\ast] =&~ (\lambda_f - \mathbb{E}[\lambda_s])\int_{\alpha r^\ast}^{\alpha r^\ast+d_\text{max}} (x-\alpha r^\ast)f(x)dx - r^\ast \nonumber\\
&~+ (\lambda_f - \mathbb{E}[\lambda_s])~d_{\text{min}}\left[ F(d_{\text{max}}) - F(d_{\text{max}}-h(r)) \right]\\
=&~ (\lambda_f - \mathbb{E}[\lambda_s]) \int_{F^{-1}(1 - \frac{1}{\alpha(\mathbb{E}[\lambda_s] - \lambda_f)})}^\infty x f(x)dx
\end{align*}

\subsection*{Proof of Theorem \ref{thm:unique_portfolio}}
This theorem can be proved by showing that the determinant of the Hessian of the two-dimensional optimization problem is negative, and hence yields a saddle at each joint minimum of portfolios ($(r^\ast, \bar{q}_C^\ast$ for DR + call, $(r^\ast, \bar{q}_F^\ast)$ for DR + forward contract, $(\bar{q}_F^\ast, \bar{q}_C^\ast)$ for call + forward contract). The objectives for each of these pairwise portfolios are
\begin{align*}
\Pi_{\text{FC}} =&~ \lambda_f d - \bar{\lambda}_F \bar{q}_F - P\bar{q}_C - (d-\bar{q}_F-\bar{q}_C)\lambda_s\mathbf{1}_{d > \bar{q}_F+\bar{q}_C} \\
&~- (d-\bar{q}_F)\min(\lambda_s, \bar{\lambda}_C)\mathbf{1}_{\bar{q}_F \leq d \leq \bar{q}_F + \bar{q}_C}\\
\Pi_{\text{FD}} =&~ (\lambda_f-\lambda_s)[d(r)-\bar{q}_F]_+ - \bar{\lambda}_F\bar{q}_F -r\\
&~+\lambda_F d(r)\mathbf{1}_{d(r) \leq \bar{q}_F} + \lambda_f\bar{q}_F\mathbf{1}_{d(r)>\bar{q}_F}\\
\Pi_{\text{CD}} =&~ \lambda_f d - \lambda_s [d(r)-\bar{q}_C]_+ -P\bar{q}_C - r\\
&+~ \min(\bar{\lambda}_C,\lambda_s)\left[-d(r)\mathbf{1}_{d(r)\leq \bar{q}_C} - \bar{q}_C\mathbf{1}_{d(r) > \bar{q}_C}\right]
\end{align*}
where $\Pi_{\text{FC}}, \Pi_{\text{FD}}$, and $\Pi_{\text{CD}}$ denote the profit under the pairwise portfolios (forward contract, call), (forward contract, DR), (call, DR), respectively. Taking the expectation w.r.t to the random variables $d$ and $\lambda_s$ and the derivatives w.r.t. the decision variables yields the Hessian matrix, from which further analysis proves the claim.

\subsection*{Proof of Proposition \ref{prop:reformulation_with_cvar}}
Using the definition of the conditional expectation for continuous random variables $X, Y$
\begin{align*}
\mathbb{E}[X|Y] = \int_{x\in\mathbb{R}}p_{X|Y}(x|y)dx,
\end{align*}
it follows that
\begin{align}\label{eq:conditional_expectation}
\mathbb{E}[d~|~d\geq \tau] = \frac{\int_\tau^{d_{\text{max}}}xf(x)dx}{\int_\tau^{d_{\text{max}}}f(x)dx},\quad d_{\text{min}} < \tau < d_{\text{max}}.
\end{align}
Applying \eqref{eq:conditional_expectation} on \eqref{eq:opt_profit_forward_contract}, \eqref{eq:opt_profit_call}, and \eqref{eq:opt_profit_DR} with $\tau = \alpha_F$ \eqref{eq:alpha_forward_contract}, $\tau = \alpha_C$ \eqref{eq:alpha_call_option}, and $\tau = \alpha_{\text{DR}}$ \eqref{eq:alpha_demand_response}, respectively, yields the desired expressions.

\subsection*{Proof of Proposition \ref{prop:reformulation_with_std}}
For a uniform distribution with support $[d_{\text{min}}, d_{\text{max}}]$, the PDF is $f(x)=1/(d_{\text{max}}-d_{\text{min}})\mathbf{1}(d_{\text{min}}\leq x\leq d_{\text{max}})$, and the inverse CDF is $F^{-1}(z) = d_{\text{min}} + (d_{\text{max}}-d_{\text{min}})z$, $z\in[0,1]$. Straightforward manipulation of the optimal expected profits \eqref{eq:opt_profit_forward_contract}, \eqref{eq:opt_profit_call}, and \eqref{eq:opt_profit_DR} and using the formula for the standard deviation
\begin{align*}
\sigma = \frac{d_{\text{max}}-d_{\text{min}}}{2\sqrt{3}}
\end{align*}
yields \eqref{eq:profit_dispersion_forward}, \eqref{eq:profit_dispersion_call}, and \eqref{eq:profit_dispersion_demand_response}.

\subsection*{Proof of Theorem \ref{thm:best_option_choice}}
Straightforward by pairwise comparison of equations \eqref{eq:opt_profit_forward_cvar}-\eqref{eq:opt_profit_dr_cvar}.

\end{document}